\documentclass[12pt,amsfonts]{iopart}

\usepackage{iopams}

\input epsf

\usepackage{graphicx}

\newcommand{\beq}{\begin{equation}}
\newcommand{\eeq}{\end{equation}}
\newcommand{\beqs}{\begin{eqnarray}}
\newcommand{\eeqs}{\end{eqnarray}}

\begin{document}

\title{Some Exact Results on the Potts Model Partition
  Function in a Magnetic Field}

\author{Shu-Chiuan Chang$^1$ and Robert Shrock$^2$}

\address{$^1$ \ Department of Physics, National Cheng Kung University, 
Tainan 70101, Taiwan}

\address{$^2$ \ C. N. Yang Institute for Theoretical Physics,
Stony Brook University, Stony Brook, NY 11794, USA}

\eads{\mailto{scchang@mail.ncku.edu.tw}, \mailto{robert.shrock@stonybrook.edu}}

\begin{abstract} 

We consider the Potts model in a magnetic field on an arbitrary graph $G$.
Using a formula of F. Y. Wu for the partition function $Z$ of this model as a
sum over spanning subgraphs of $G$, we prove some properties of $Z$ concerning
factorization, monotonicity, and zeros. A generalization of the Tutte
polynomial is presented that corresponds to this partition function.  In this
context we formulate and discuss two weighted graph-coloring problems.  We also
give a general structural result for $Z$ for cyclic strip graphs.

\end{abstract} 



\maketitle

\pagestyle{plain}
\pagenumbering{arabic}
\setcounter{equation}{0}

The $q$-state Potts model has served as a valuable system for the study of
phase transitions and critical phenomena \cite{potts}-\cite{baxterbook} and
has interesting connections with mathematical graph theory
\cite{welsh}-\cite{jemrev}. On a lattice, or, more generally, on a graph $G$,
at temperature $T$ and in an external magnetic field $H$, this model is defined
by the partition function
\beq
Z = \sum_{ \{ \sigma_i \} } e^{-\beta {\cal H}}
\label{z}
\eeq
 with the Hamiltonian
\beq
{\cal H} = - \sum_{\langle i j \rangle} J_{ij} \delta_{\sigma_i, \sigma_j}
- H \sum_i \delta_{\sigma_i, 1} \ , 
\label{ham}
\eeq
where $\sigma_i=1,...,q$ are classical spin variables on each vertex (site) $i
\in G$; $\beta = (k_BT)^{-1}$, $\langle i j \rangle$ denote pairs of adjacent
vertices, and $J_{ij}$ are the associated spin-spin couplings.  The graph
$G=G(V,E)$ is defined by its vertex set $V$ and its edge (bond) set $E$;
we denote the number of vertices of $G$ as $n=n(G)$ and the number of edges of
$G$ as $e(G)$. With no loss of generality, we take $G$ to be connected and the
external field to pick out the value $\sigma_i=1$ from the $q$
possible values. We first consider the case of a single spin-spin coupling
$J_{ij}=J$ and use the notation
\beq
K = \beta J \ , \quad h = \beta H , \quad y=e^K, \quad
v = y - 1 \ , \quad w=e^h \ . 
\label{kdef}
\eeq
From (\ref{clusterw}) below, it follows that $Z$ is a polynomial in $q$, $v$,
and $w$, so we write $Z=Z(G,q,v,w)$ and, for the zero-field ($w=1$) case, we
set $Z(G,q,v) \equiv Z(G,q,v,1)$.  Positive $H$ gives a weighting that favors
spin configurations in which $\sigma_i$'s have the value 1, while negative $H$
disfavors such configurations.  In the limit $h \to -\infty$, configurations in
which any $\sigma_i=1$ make no contribution to $Z$, so that the model reduces
to the zero-field case with $q$ replaced by $q-1$:
\beq
Z(G,q,v,0) = Z(G,q-1,v,1) \ . 
\label{zw10}
\eeq

The original definition of the Potts model, (\ref{z}) and (\ref{ham}), 
requires $q$ to be a positive integer, $q \in {\mathbb N}_+$.  This 
restriction is removed for the zero-field Potts model by the Fortuin-Kasteleyn
representation 
\cite{fk}
\beq
Z(G,q,v) = \sum_{G' \subseteq G} v^{e(G')} q^{k(G')}  \ , 
\label{cluster}
\eeq
where $G'=(V,E')$, $E' \subseteq E$ is a spanning subgraph of $G$, and
$k(G')$ denotes the number of (connected) components of $G'$. Eq.
(\ref{cluster}) has the crucial property that $Z(G,q,v)$ is expressed in a
manner that does not make any explicit reference to the spins $\{\sigma_i\}$ or
summation over spin configurations, $\sum_{\{\sigma_i\}}$. This enables one to
define the zero-field model Potts model partition function for any real $q \ge
0$. For the ferromagnetic case, $v > 0$, so $Z(G,q,v) > 0$ for $q > 0$ and
hence (\ref{cluster}) defines a Gibbs measure.  For the antiferromagnetic case,
since $v$ is negative ($-1 \le v \le 0$), (\ref{cluster}) does not, in general,
yield a positive-definite $Z$ with Gibbs measure if $q \not\in {\mathbb
N}_+$. Eq.  (\ref{cluster}) also establishes the equivalence of the zero-field
Potts partition function to the Tutte polynomial $T(G,x,y)$, a function of
major importance in graph theory, 
\beq
T(G,x,y) = \sum_{G' \subseteq G} (x-1)^{k(G')-k(G)}(y-1)^{c(G')} \ , 
\label{t}
\eeq
where $c(G')=e(G')+k(G')-n(G')$ is the number of independent cycles on $G'$
\cite{welsh}-\cite{jemrev},\cite{tutte}-\cite{boll}. The equivalence is
$Z(G,q,v) = (x-1)^{k(G)}(y-1)^{n(G)}T(G,x,y)$, where $x=1+(q/v)$.

The Fortuin-Kasteleyn cluster formula (\ref{cluster}) was generalized to the
case of a nonzero external magnetic field by F. Y. Wu \cite{wurev,wu78}.
Denote each of the connected components of $G'$ as $G'_i$, $i=1,...,k(G')$.
Wu's result is \cite{wurev,wu78} 
\beq
Z(G,q,v,w) = \sum_{G' \subseteq G} v^{e(G')} \
\prod_{i=1}^{k(G')} \Big ( q-1 + w^{n(G'_i)} \Big ) \ . 
\label{clusterw}
\eeq
We first use the Wu formula (\ref{clusterw}) to prove a number of properties of
$Z(G,q,v,w)$ concerning factorization, monotonicity, and zeros. Combining
(\ref{cluster}), which shows that $Z(G,q,v)$ contains a factor of $q$, with
(\ref{zw10}), we deduce that $Z(G,q,v,0)$ contains a factor of $(q-1)$.
Substituting $q=0$ in (\ref{clusterw}) and using the factorization
$w^{n(G'_i)}-1 = \tilde w \sum_{\ell=0}^{n(G'_i)-1} (\tilde w+1)^\ell$, where
$\tilde w = w-1$, we prove that $Z(G,0,v,w)$ contains a factor of $(w-1)$.
Setting $w=q-1$ in (\ref{clusterw}) yields the result that $Z(G,q,v,q-1)$ has
$(q-1)$ as a factor. Substituting $w=0$ in (\ref{clusterw}) is another way to
derive (\ref{zw10}).  Two elementary results are
$Z(G,1,v,w)=(v+1)^{e(G)}w^{n(G)}$ and $Z(G,q,0,w)=(q-1+w)^{n(G)}$.

We can write $Z(G,q,v,w)$ in several equivalent ways:
\beqs
& & Z(G,q,v,w) = \sum_{r,t=0}^{n(G)} \, \sum_{s=0}^{e(G)} \ a_{rst} \, 
q^r v^s w^t  \ = \
\sum_{r,t=0}^{n(G)} \, \sum_{s=0}^{e(G)} \ b_{rst} \, q^r y^s w^t  \cr\cr
& = & 
\sum_{r,t=0}^{n(G)} \, \sum_{s=0}^{e(G)} \ c_{rst} \, 
\tilde q^{\, r} v^s w^t \ = \
\sum_{r,t=0}^{n(G)} \, \sum_{s=0}^{e(G)} \ d_{rst} \, q^r v^s \tilde w^{\, t}
 \ , 
\label{zgenformwv}
\eeqs
where $\tilde w = w-1$ as before, $\tilde q = q-1$, and the coefficients
$a_{rst}$, $b_{rst}$, $c_{rst}$, and $d_{rst}$ are integers. Some $a_{rst}$ and
$b_{rst}$ can be negative. In contrast, the Wu formula (\ref{clusterw}) shows
that all of the nonzero $c_{rst}$ are positive. This leads to three
monotonicity and zero-free properties in the corresponding variables $\tilde
q$, $v$, and $w$, taken here as real: (i) for $\tilde q > 0$ and $v > 0$,
$Z(G,q,v,w) \equiv Z$ is a monotonically increasing function (MIF) of $w > 0$
and has no zeros on the positive $w$ axis; (ii) for $v > 0$ and $w > 0$, $Z$ is
a MIF of $\tilde q > 0$ and has no zeros on the positive $\tilde q$ axis; (iii)
for $w > 0$ and $\tilde q > 0$, $Z$ is a MIF of $v > 0$ and has no zeros on the
positive $v$ axis.  We can also prove that all of the nonzero $d_{rst}$ are
positive by using (\ref{clusterw}) together with the relation used above,
$w^{n(G'_i)}-1= \tilde w \sum_{\ell=0}^{n(G'_i)-1} (\tilde w+1)^\ell$.  Since
each term in the expansion of $(\tilde w+1)^\ell$ is positive for each $G'_i$,
this shows that the nonzero $d_{rst}$ are positive.  This yields three more
monotonicity and zero-free results (which have some overlap with (i)-(iii)):
(iv) for $q > 0$ and $v > 0$, $Z$ is a MIF of $\tilde w > 0$ and has no zeros
on the positive $\tilde w$ axis; (v) for $v > 0$ and $\tilde w > 0$, $Z$ is a
MIF of $q > 0$ and has no zeros on the positive $q$ axis; (vi) for $\tilde w >
0$ and $q > 0$, $Z$ is a MIF of $v > 0$ and has no zeros on the positive $v$
axis.  Monotonicity relations for borderline cases are covered by our results
above; e.g. for $q=1$, $Z(G,1,v,w)$ is a MIF of $v > -1$ for $w > 0$ and a MIF
of $w > 0$ for $v > -1$; for $v=0$, $Z(G,q,0,w)$ is a MIF of $q-1+w > 0$, etc.

We define a rational function that generalizes the Tutte polynomial, namely 
\beqs
U(G,x,y,w) & = & (x-1)^{-k(G)}(y-1)^{-n(G)}\sum_{G' \subseteq G} (y-1)^{e(G')} 
\ \times \cr\cr
& \times & \prod_{i=1}^{k(G')} (xy-x-y+w^{n(G'_i)}) \ . 
\label{u}
\eeqs
This function satisfies $U(G,x,y,w)=(x-1)^{-k(G)}(y-1)^{-n(G)}Z(G,q,v,w)$ and 
$U(G,x,y,1)=T(G,x,y)$. Although $T(G,x,y)$ and
$Z(G,q,v)$ satisfy deletion-contraction relations, we note that for $w$ not
equal to 1 or 0, the functions $U(G,x,y,w)$ and $Z(G,q,v,w)$ do not, in
general, satisfy such deletion-contraction relations.

We define two types of graph coloring problems and use special cases of
(\ref{clusterw}) to describe these.  Although graph coloring has been
investigated intensively \cite{welsh}-\cite{jemrev}, \cite{biggs}-\cite{boll},
\cite{jensentoft}, these two types of graph colorings have not, to our
knowledge, been studied before. Recall that the chromatic polynomial $P(G,q)$
counts the number of ways of assigning $q$ colors to the vertices of a graph
$G$ such that no adjacent vertices have the same color.  This ``proper
$q$-coloring'' of the vertices of $G$ is equivalent to $Z$ for the
zero-temperature, zero-field Potts antiferromagnet, $v=-1$: $P(G,q)=Z(G,q,-1)$.
We generalize this to a weighted proper $q$-coloring of the vertices of $G$, as
described by the polynomial $Ph(G,q,w) = Z(G,q,-1,w)$.  For $H < 0$, i.e., $0
\le w < 1$, we have a weighted graph coloring problem in which one carries out
a proper $q$-coloring of the vertices of $G$ but with a penalty factor of
$w$ for each vertex assigned the color 1.  For $H > 0$, we have a second type
of weighted graph coloring problem, namely a proper vertex $q$-coloring with a
weighting that favors one color.  Since this favoring of one color conflicts
with the constraint that no two adjacent vertices have the same color,
the range $w > 1$ involves competing interactions and frustration.  

Both of these weighted graph coloring problems have physical applications.  For
example, the weighted coloring problem with $0 < w < 1$ describes the
assignment of frequencies to commercial radio broadcasting stations in an area
such that (i) adjacent stations must use different frequencies to avoid
interference and (ii) stations prefer to avoid transmitting on one particular
frequency, e.g., because it is used for data-taking by a nearby radio astronomy
antenna. The graph coloring problem with $w > 1$ describes this frequency
assignment process with a preference for one of the $q$ frequencies, e.g.,
because it is most free of interference. We note some other special cases. Just
as the Tutte polynomial $T(G,1-q,0)$ gives, up to a prefactor, $P(G,q)$, so
also $T(G,0,1-q)$ determines the flow polynomial $F(G,q)$, which counts the
number of nowhere-zero $q$-flows on $G$ that satisfy flow conservation mod $q$
at each vertex.  The function $U(G,0,1-q,w)$ then defines a weighted flow
problem.  With $0 < w < 1$, this could describe a discretized flow analysis in
an electrical circuit or traffic flow situation in which one incorporates a
finite penalization for one, say the maximal, flow, in order to minimize power
dissipation in resistors in the circuit case or to minimize traffic jams in the
traffic case.

For a planar $G$, $P(G,q)$ counts not just the number of proper $q$-colorings
of $G$ vertices but also, equivalently, the number of proper $q$-colorings of
the faces of the dual graph $G^*$. Similarly, for planar $G$, $Ph(G,q,w)$ is a
measure not only of the weighted proper $q$-colorings of $G$ vertices, but
also, equivalently, the weighted proper $q$-colorings $G^*$ faces.

We have used (\ref{clusterw}) and combinatoric arguments of the type in
\cite{cf} to obtain a general structural determination of $Z(G,q,v,w)$ for
cyclic and M\"obius strip graphs $G_s$ of regular lattices of fixed width $L_y$
vertices and arbitrary length as well as self-dual strips of the square
lattice, extending \cite{dg}. This length is $L_x \equiv m$ ($L_x \equiv 2m$)
for square and triangular (honeycomb) strips. For cyclic $G_s$ we find
\beq
Z(G_s,q,v,w) = \sum_{d=0}^{L_y} \sum_{j=1}^{n_{Zh}(L_y,d)} \tilde c^{(d)}(q) 
[\lambda_{G_s,L_y,d,j}(q,v,w)]^m \ , 
\label{zgsumw}
\eeq
where $Zh$ connotes $Z$ for $h \ne 0$ and 
\beq
\tilde c^{(d)} = \sum_{j=0}^d (-1)^j {2d-j \choose j}(q-1)^{d-j} \ . 
\label{barc}
\eeq
We have $n_{Zh}(L_y,L_y)=1$, $n_{Zh}(1,0)=2$, and $n_{Zh}(L_y,d)=0$ for $d >
L_y$; the other $n_{Zh}(L_y,d)$ are determined by the recursion relations
$n_{Zh}(L_y+1,0) = 2n_{Zh}(L_y,0) + n_{Zh}(L_y,1)$ and, for $1 \le d \le
L_y+1$,
\beq
n_{Zh}(L_y+1,d) = n_{Zh}(L_y,d-1) + 3n_{Zh}(L_y,d) + n_{Zh}(L_y,d+1) \ . 
\label{ntrecursion2}
\eeq
The form for M\"obius strips involves switches of certain $\tilde c^{(d)}$'s
(generalizing switchings in the $w=1$ case \cite{cf}), which are given in
detail elsewhere \cite{zth}. For these cyclic (and M\"obius) strip graphs of
width $L_y$, the total number of different $\lambda$'s,
$N_{Zh,L_y}=\sum_{d=0}^{L_y} n_{Zh}(L_y,d)$, is
\beq
N_{Zh,L_y} = \sum_{j=0}^{L_y} {L_y \choose j} {2j \choose j} \ . 
\eeq

It is straightforward to generalize (\ref{clusterw}) to the case where the
spin-spin couplings $J_{ij}$ depend on the edges $e_{ij}$. Let us define 
$K_{ij}=\beta J_{ij}$, $y_{ij}=e^{K_{ij}}$, $v_{ij}=y_{ij}-1$, and 
the set of $v_e$ for $e \equiv e_{ij} \in E$ as $\{v_e\}$. Then we have 
\beq
Z(G,q,\{v_e\},w) = \sum_{G' \subseteq G} \Big [ \prod_{e \in E'} v_e \Big ] \
\Big [ \prod_{i=1}^{k(G')} ( q-1 + w^{n(G'_i)} ) \Big ] \ . 
\label{clusterwv}
\eeq

We thank F. Y. Wu for a valuable communication calling our attention to Ref. 
\cite{wu78}.  This research was partly supported by the grants 
NSC-97-2112-M-006-007-MY3, NSC-98-2119-M-002-001 (S.-C.C.), and 
NSF-PHY-06-53342 (R.S.).

\end{document}